\documentstyle [11pt,epsf] {article}

\begin{document}
\begin{titlepage}

\begin{center}
\vspace{2cm}
\LARGE
Feast and Famine: Regulation of Black Hole Growth in Low Redshift Galaxies    
\\                                                     
\large
\vspace {0.6cm}
Guinevere Kauffmann$^1$ and Timothy M. Heckman$^2$, 
\\
\end{center}
\small
{\em $^1$Max-Planck Institut f\"{u}r Astrophysik, D-85748 Garching, Germany} \\
{\em $^2$ Center for Astrophysical Sciences, Department of Physics and Astronomy, Johns Hopkins University, Baltimore, MD 21218}\\
\Large           
\begin {abstract}
\normalsize
We analyze the observed distribution of Eddington ratios ($L/L_{Edd}$) as a function of supermassive black hole mass for a large sample of nearby galaxies drawn from the Sloan Digital Sky Survey. We demonstrate that there are two distinct regimes of black hole growth in nearby galaxies. 
The first  is associated with galaxies with significant star formation ($M_*/SFR \sim$ a Hubble time) in
their central kiloparsec regions, and is characterized by a broad log-normal distribution of accretion rates peaked 
at a few percent of the Eddington limit. In this regime, the Eddington ratio distribution is independent 
of the mass of the black hole and shows little dependence on the central stellar population of the galaxy. 
The second regime is associated with galaxies with old central stellar populations ($M_*/SFR >>$ a Hubble time), and is characterized by a power-law distribution function of Eddington ratios. In this regime, the time-averaged mass accretion rate onto black holes is proportional to the mass of stars in the galaxy bulge, with a constant of proportionality that depends on the mean stellar age of the stars. This result is once again independent of black hole mass.  We show that both the slope of the power-law and
the decrease in the accretion rate onto black holes in old galaxies are consistent with population synthesis model predictions of the decline in stellar mass loss rates as a function of mean stellar age. Our results lead to a very simple picture of  black hole growth in the local Universe. If the supply of cold gas in a galaxy bulge is plentiful, the black hole regulates its own growth at a rate that does not further depend on the properties of the interstellar medium. Once the gas runs out, black hole growth is regulated by the rate at which evolved stars lose their mass.   
\end {abstract}
\vspace {0.8 cm}
\end {titlepage}
\normalsize

\section {Introduction}

Large surveys of  active galactic nuclei (AGN)  teach us a great deal about the growth of super-massive black holes in galaxies as a function of cosmic time. By comparing quasar counts at different redshifts to the mass and  number densities of black holes in the present-day Universe, one can derive constraints on the radiative efficiencies and lifetimes of these systems (e.g. Soltan 1982; Yu \& Tremaine 2002). Additional constraints on quasar lifetimes also come from the measured clustering amplitude of these objects (e.g. Martini \& Weinberg 2001; Shen et al 2007; Shankar et al 2008). There is now general consensus that optically luminous quasars are radiating close to their Eddington limit
(Kollmeier et al. 2006), have short($\sim 10^7$ yr) lifetimes, and in total account for a sizeable fraction of the total mass density in black holes at the present day (Yu \& Tremaine 2002).  

By number, however, quasars only account for a small fraction of the total AGN population. Studies of complete samples of AGN in the local Universe (e.g. Heckman 1980, Ho, Filippenko \& Sargent 1997; Kauffmann et al 2003b; Hao et al 2005) have taught us that the AGN fraction in massive galaxies 
with $> 10^{10.5} M_{\odot}$ galaxies is very high ($>$ 50 \%) and that AGN span a large range in Eddington ratios (Heckman et al 2004; Ho 2008). In lower-luminosity and optically-obscured (Type 2) AGN, the active nucleus does not outshine the underlying host galaxy, so one can carry out joint analyses of the accretion rate as a function of the physical properties of the host, and attempt to figure out which processes may be involved in triggering  accretion onto the central black hole. Such studies have demonstrated that black hole growth is generally more rapid in low-redshift galaxies with younger stellar populations (Kauffmann et al. 2003b; Cid Fernandes et al. 2004). Moreover, the average ratio between the mass of stars forming in the bulge and the mass being accreted by the black hole is $\sim 1000$ (Heckman et al. 2004), which is very similar to the ratio between bulge mass and black hole mass in present-day inactive elliptical galaxies and bulges (Marconi \& Hunt 2003; Haring \& Rix 2004). This provides important confirmation that bulge formation and black hole growth are indeed still connected today, at least in an average sense.

It is important to try and pinpoint which {\em physical mechanisms} are responsible for the link between black hole growth and star formation in the host galaxy. One idea that has gained considerable popularity in recent years is
that gravitational torques that arise during galaxy-galaxy mergers or interactions drive gas into the nuclear regions of galaxies, where it is able
to fuel a central starburst and also accrete onto the black hole (e.g.
Sanders et al 1988; Kauffmann \& Haehnelt 2000; Granato et al 2004; Croton et al 2006). It is possible to model these processes in considerable detail using numerical simulations (Barnes \& Hernquist 1991,1996; Mihos \& Hernquist 1996). 
In more recent simulations (Di Matteo et al 2005; Springel et al 2005), some of the energy of the accreting black holes couples to the surrounding gas, and black hole growth eventually shuts down when the feedback energy is sufficient to unbind the reservoir of gas. Hopkins et al (2005 a,b) have used these simulations to derive physically-motivated light curves for AGN as a function of time in a large suite of simulations of merging galaxies, and to derive from this the luminosity-dependent lifetime distribution of AGN, under the assumption that they are triggered by mergers. In a more recent paper, Hopkins \& Hernquist (2008) show how  observed Eddington ratio distributions for  complete samples of AGN are able to place strong constraints on different models for how black holes are fueled, and find  that the available data agrees with the merger model predictions.

Observationally, there is still considerable debate as to whether any link
exists between mergers/interactions and accretion onto the black hole. The best
current constraints again come from studies of large samples of low redshift AGN. Recent clustering studies demonstrate that the star formation rates in galaxies are strongly enhanced if they have close neighbors, but that there is no similar enhancement in nuclear activity (Li et al 2006, 2008). Similarly, Reichard et al (2008) have used the 'lopsidedness' in the stellar light distribution of SDSS galaxies as a signpost of interactions and mergers. They find a strong link between lopsidedness and star formation, but they also demonstrate that if AGN and non-AGN samples are matched to have the same stellar
masses, structural properties and central stellar populations, there is no
difference in the lopsidedness of the light distribution between the active
and non-active galaxies. If mergers and interactions are not instrumental 
in fueling accretion onto the central black holes of nearby galaxies, then other mechanisms such as bar-driven instabilities (Shlosman, Frank \& Begelman 1989; Garcia-Burillo et al 2005) or stellar mass loss in galactic bulges
(Mathews \& Baker 1971; Norman \& Scoville 1988; Ciotti \& Ostriker 1997,2007) may play a more important role in explaining observed connection between AGN activity and recent star formation in the bulge. 

In this paper, we re-examine the relationship between the accretion rates onto black holes and star formation in their host galaxies. Samples of nearby AGN are now sufficiently large to enable us to study not only the average relation between black hole accretion and star formation, but also how the {\em distribution functions} of black hole accretion rates vary as a function of black hole mass, and how the shapes of these distribution functions are modulated as the physical conditions and stellar populations in the galaxy change. As we will show, the full distribution function of accretion rates does indeed contain new information, and allows us to identify two clearly distinct "modes" of black hole growth in the local Universe. One mode is connected with galaxies with an ample supply of cold gas, and the other is connected with galaxies with aging and passively evolving stellar populations, where cold gas is presumably in much shorter supply. We then speculate on the physical origin of the two modes in view of recent ideas about how black holes are formed and fueled.

\section{Methodology}

We are interested in this paper in calculating the distribution of the Eddington ratio ($L_{bol}/L_{Edd}$) as a function of the key properties of the galaxies and their black holes. Our sample is drawn from the Data Release 4 (DR4) of the Sloan Digital Sky Survey (York et al 2000; Adelman-McCarthey et al 2006).
The spectroscopic quantities we refer to are available for public download at
http://www.mpa-garching.mpg.de/SDSS/DR4/.

To examine the dependence of the distribution of the Eddington ratio on the star formation history of the galaxy, we will use the amplitude of the 4000 \AA\ break (as measured in the SDSS spectra with the D$_n$(4000) index introduced
by Balogh et al. (1999)). As discussed by Kauffmann et al. (2003a), 
the value of D$_n$(4000) increases monotonically as the luminosity-weighted mean age of the stellar population increases. This dependence was quantified empirically by Brinchmann et al. (2004) who used SDSS spectroscopic measurements of the luminosity of the extinction-corrected H$\alpha$ emission-line in star forming galaxies to relate the star formation rate per unit stellar mass (SFR/M$_*$) to the value of D$_n$(4000). Obviously these two quantities are only related in an average sense, because 
the Brinchmann et al measurements were derived using emission line fluxes, which come from stars with ages less than
$10^7$ years, whereas   D$_n$(4000) is sensitive to young stars with a much broader range in age ($>1$ Gyr).   

We will estimate the black hole mass using the stellar velocity dispersions measured from the SDSS spectra and the formula given in Tremaine et al (2002). 
To estimate the bolometric luminosity, we will use the luminosity of the [OIII]$\lambda$5007 
emission-line. The rationale for this is discussed in some detail in Heckman et al. (2004), but the primary reasons are that the [OIII] emission-line is one of the strongest lines in SDSS spectra and that it suffers substantially less contamination by emission from star forming regions than other strong lines. We will return to this latter point below. In Heckman et al (2004) the [OIII] line luminosities were not corrected for extinction by dust. This was because we used calibrations between [OIII] line luminosity and bolometric luminosity for Type 1 Seyferts and low-z quasars to calibrate L[OIII]/$M_{BH}$ in units of the Eddington luminosity. Balmer decrement measurements for the narrow emission-lines are not generally available for the Type 1 Seyferts and quasars. We thus decided to work with uncorrected line luminosities. The Heckman et al (2004) paper was mainly concerned with {\em average} accretion rates. In this paper we will be determining the distributions of the accretion rate as a function of the age of the stellar population. Dust corrections can be substantial, and will be systematically higher for AGN in galaxies with higher levels of ongoing star formation (Kauffmann et al. 2003b). Thus in this paper we will apply dust extinction corrections using the measured Balmer decrements in conjunction with
a standard extinction law ( following the procedure outlined in Kewley et al (2002).

While we will present our results in terms of the luminosity ratio log L[OIII]/M$_{BH}$ (hereafter, the Eddington parameter), it is important to relate this to the actual Eddington ratio ($L_{bol}/L_{Edd}$). To do so, we need a bolometric correction to the extinction corrected [OIII] luminosity. In Heckman et al. (2004) we showed that for Type 1 Seyfert nuclei and low-z quasars, the average bolometric correction to the {\it uncorrected} [OIII] luminosity was 3500. Typical measured values for the Balmer decrement in powerful AGN imply mean extinction corrections of $\sim$1.5 to 2 magnitudes (e.g Kewley et al. 2006), so we would expect the typical bolometric correction to the extinction corrected [OIII] luminosity for such AGN to be in the range $\sim$ 600 to 800. A more direct estimate comes from the combination of XMM X-ray, SDSS optical, and Spitzer mid-IR spectra of a complete flux-limited sample of SDSS Type 2 AGN (Heckman et al. in preparation). These data imply corresponding bolometric corrections in the range $\sim$500 to 800.

These values pertain to moderately powerful AGN with log L[OIII]/M$_{BH} >$ -0.5 (typically Type 2 Seyfert nuclei). The situation is less clear for the lower luminosity AGN (typically LINERs). First of all, the overall ionization state of the emission-line region is lower than in Seyfert nuclei, and so the [OIII] line will carry a smaller fraction of the total ionizing luminosity (Netzer 2009,
in preparation). Second, the ratio of ionizing to total luminosity may be significantly different compared to the more powerful AGN (Ho 2008, but see Maoz 2007). We have therefore considered the low-luminosity AGN separately. Ho (2008) has derived a mean bolometric correction to the extinction-corrected H$\alpha$ luminosity of $\sim$220 for low-luminosity AGN. Compared to the [OIII] line, the H$\alpha$ emission-line suffers much more severe contamination from regions of star formation encompassed by the SDSS fiber (as quantified below). We have therefore converted the bolometric correction in Ho (2008) to the corresponding value for the [OIII] luminosity of $\sim$ 300 to 600. Since the ranges in the bolometric corrections for the lower- and higher-powered AGN overlap, for simplicity we will simply take a mean bolometric correction to the extinction-corrected [OIII] of 600. Thus, for $L/L_{Edd} = 1$, the corresponding value of log L[OIII]/M$_{BH}$ would be $\sim$1.7 dex. 

In our past work, AGN have been identified according to their location in the
Baldwin, Philipps \& Terlevich (BPT; 1981) diagrams. In particular, in the space of [OIII]$\lambda$ 5007/H$\beta$ versus  [NII]$\lambda$6583/H$\alpha$, AGN and star-forming galaxies are relatively well separated, with the AGN emerging
as a plume from the bottom of the star-forming sequence, and extending upwards and outwards towards larger [OIII]/H$\beta$ and [NII]/H$\alpha$ values. Kauffmann et al. (2003b) used data from the SDSS to define a fiducial cut to 
separate star-forming galaxies from AGN. This is illustrated in Figure 1
where we plot BPT diagrams for four volume-limited samples of galaxies with $0.02 < z< 0.1$ divided according to black hole mass. We note that the SDSS spectra are measured using 3 arc second diameter fibres, which corresponds to a physical radii of $\sim 0.6 -2.7$ kpc 
for the AGN in our sample (i.e. the spectra are mainly sampling the light from the bulges of the galaxies). 
In Figure 1, the dashed red curve shows our fiducial division between star-forming galaxies and AGN. AGN are plotted as red dots, while star-forming galaxies are plotted as black dots. We have included all the galaxies with $S/N>3$ in each of the four emission lines used to define the BPT diagram. 

\begin{figure}
\centerline{
\epsfxsize=13cm \epsfbox{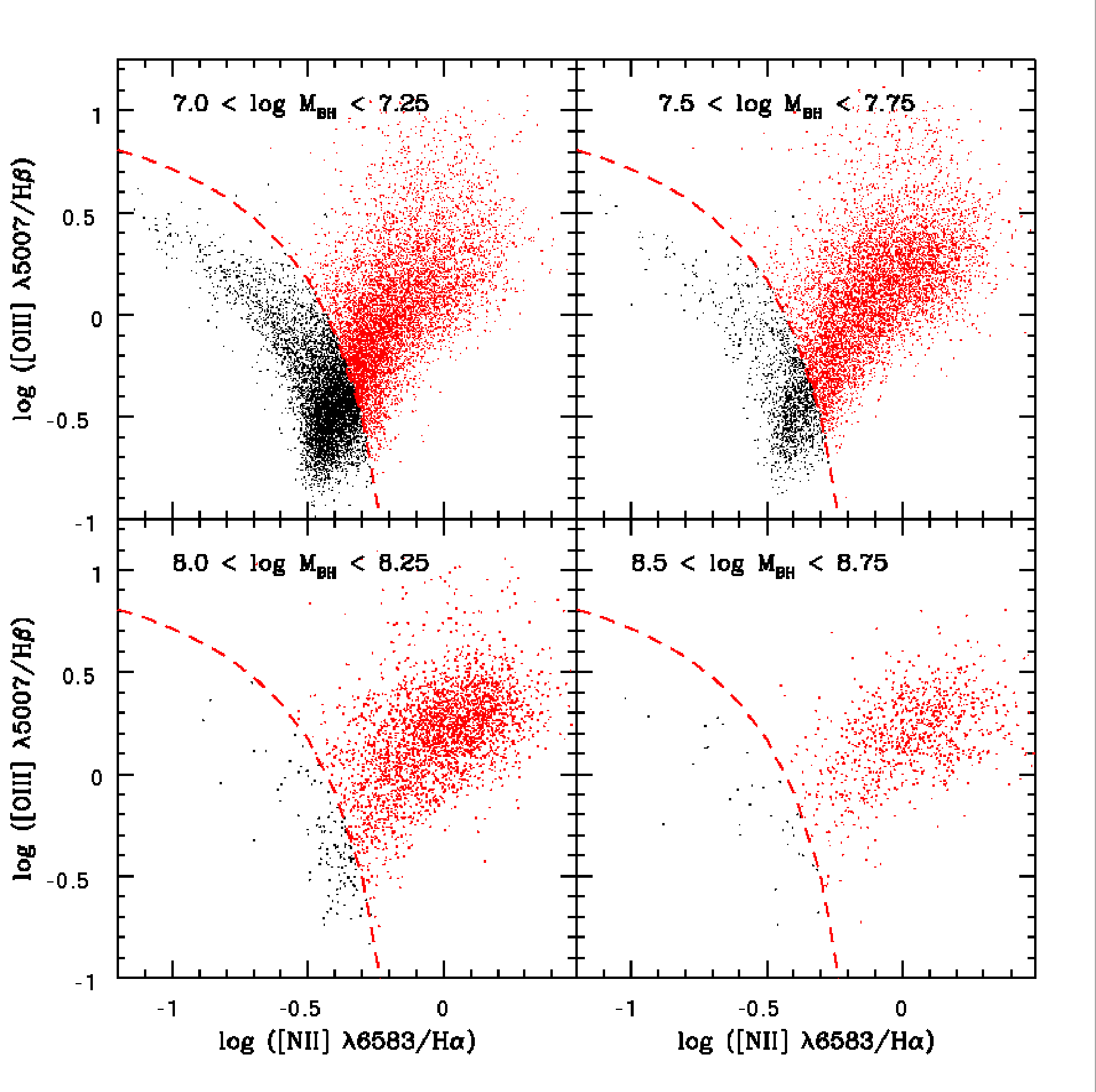}
}
\caption{\label{fig1}
\small
[OIII]$\lambda$ 5007/H$\beta$ versus [NII]$\lambda$6583/H$\alpha$
BPT diagrams  for a volume-limited sample of galaxies with $z<0.1$. Only galaxies with $S/N>3$ measurements of all four lines are shown. Results are shown in 4 different ranges of black hole mass. The dashed curve is the fiducial separation curve between star-forming galaxies and AGN defined in Kauffmann et al (2003b). Galaxies classified as AGN are plotted as red points, while galaxies classified as star-forming galaxies are plotted as black points.}
\end {figure}
\normalsize

\begin{figure}
\centerline{
\epsfxsize=13cm \epsfbox{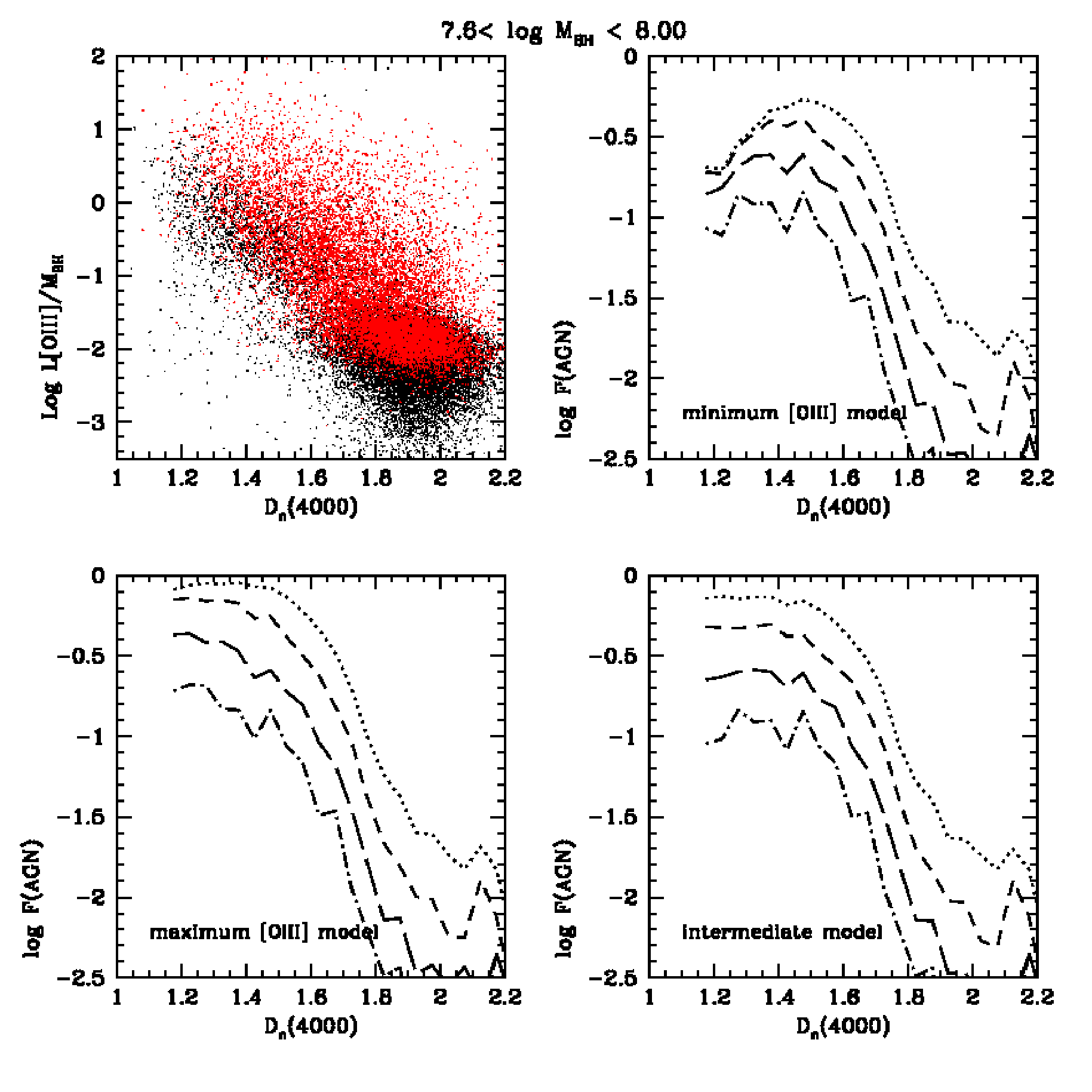}
}
\caption{\label{fig2}
\small
{\it Top left:} L[OIII]/M$_{BH}$ is plotted as a function of
4000 \AA\ break strength for a volume-limited  sample of galaxies with
$z<0.1$ and black hole masses in the range
$7.6< \log M_{BH} < 8.0$.
All galaxies are included in this diagram: if the [OIII] line is
not detected, the upper limit in L[OIII]/M$_{BH}$ is plotted.
Black points denote star-forming galaxies  or galaxies
with emission lines that are too weak to classify.  
Red points denote AGN (the classification of AGN is the same as in Figure 1).
{\it Top right:} The fraction of galaxies with  L[OIII]/M$_{BH}$   greater
than a given value is plotted as a function of D$_n$(4000). In this panel,
L[OIII] is set to zero for all galaxies classified as star-forming on the BPT
diagram. Dotted, short-dashed, long-dashed and dashed-dotted curves show results
for  L[OIII]/M$_{BH}$ =-1.0, -0.5, 0.0 and 0.4, respectively.
{\it Bottom left:} In this panel, L[OIII] is set to its measured
value  for  galaxies classified as star-forming.
{\it Bottom right:} In this panel, L[OIII] is set to one-third its measured
value  for  galaxies classified as star-forming.}
\end {figure}
\normalsize

\begin{figure}
\centerline{
\epsfxsize=12cm \epsfbox{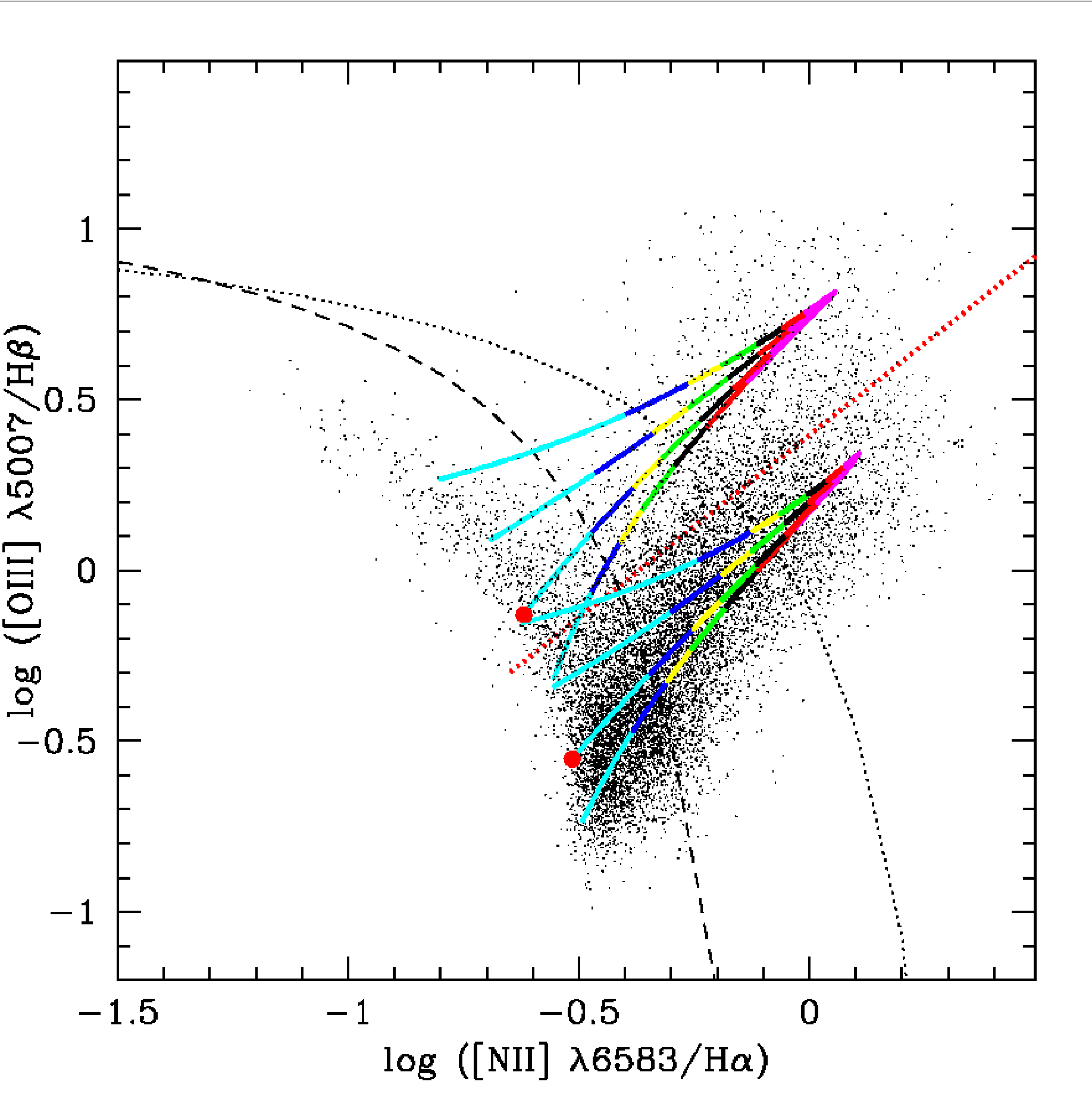}
}
\caption{\label{fig3}
An illustration of star-forming/AGN mixing trajectories. Each curve
shown on the plot connects a point on the star-forming locus to
two possible ``pure'' AGN. The curves are coloured according
to the fraction of the total [OIII] luminosity that originates from 
star formation. Cyan shows the part of the trajectory where the contribution
from star formation is greater than a half, blue  is where it is between
1/3 and 1/2, yellow between 1/4 and 1/3, green between 1/6 and 1/4,
black between 1/6 and 1/9,  red  between 1/9 and 1/16
and magenta is where the SF contribution is less than 1/16th of the total
[OIII] luminosity. The black dashed line shows the separation between star-forming
galaxies and AGN 
defined by Kauffmann et al (2003b), while the black dotted line is the 
maximal starburst demarcation of Kewley et al (2001). The two red circles
indicate the two trajectories that we use to correct the [OIII]
luminosities of the galaxies in our sample, depending on whether they lie 
above or below the red dotted line (see text for more details).}
\small
\end {figure}
\normalsize

There are two important points to note from Figure 1. First, there are a significant number of galaxies that lie near the boundary between star-forming galaxies and AGN, i.e. the separation between the two classes of objects is not a clean one. In fact, for the lower mass black holes (log $M_{BH} <$ 7.75), the density of plotted points is actually constant or even rising as we approach the boundary and cross over from objects classified as AGN to those classified as star forming. This demonstrates that a sharp boundary dividing AGN and star forming galaxies is unphysical. The data simply define a continuum along which the relative contribution of the AGN and star-forming regions varies smoothly.
Second, a larger fraction of galaxies with low mass black holes fall below our fiducial curve and are classified as star-forming galaxies. There are two reasons for this: 1) galaxies with low mass black holes have more ongoing star formation. For a fixed black hole accretion rate, the net ionizing radiation
field will be softer, causing galaxies to fall closer to the star-forming locus,
2) galaxies with low mass black holes are smaller, so the 3 arc second fibre 
aperture will sample light than originates not only from the bulge of the galaxy, but also from the star-forming disk. Kauffmann et al (2003b) show that the age of the stellar population increases as a function of the distance of the AGN from the star-forming locus, which is consistent with both these hypotheses. 

In Figure 2 (upper left) we plot our Eddington parameter, as a function of 4000 \AA\ break strength, for a complete, volume-limited sample of galaxies with black hole masses in the range $10^{7.6}$ to $10^8 M_{\odot}$. Galaxies classified as AGN are plotted as red points and galaxies classified as either star-forming galaxies or galaxies with undetected [OIII] emission-lines are plotted as black points (in the latter case the points are plotted at the value of the upper limit to the [OIII] luminosity). 

This figure is useful to address two possible limitations of our data. The first is the fact that we will no longer be able to detect the [OIII] emission-line if it is too weak relative to the underlying stellar continuum. In fact, the abrupt disappearance of AGN below log L[OIII]/M$_{BH} \sim$ -2.5 for large values of D$_n(4000)$ simply reflects this limit. In the rest of the paper we will restrict our analysis to systems above this value. The second limitation has already been mentioned above (contamination of the emission-lines by star forming regions can mask the presence of an AGN). In this regard it is reassuring that we see in Figure 2 that for the younger bulges, the numbers of AGN simply decline smoothly towards low luminosities. In fact, there are rather few galaxies  in the lower left quadrant of this figure. Thus, the decline in the number of AGN towards low luminosity in the younger bulges can not be produced by the misclassification of AGN as pure star forming galaxies. 

While contamination by regions of star formation will not have a strong qualitative impact, it does pose a problem if we wish to derive a {\em complete} and accurate distribution function of Eddington ratios and understand how the distribution is modulated by how much star formation there is in the galaxy.
That is, even in galaxies classified as star-forming, some fraction of the measured [OIII] line luminosity should probably be attributed to the AGN. We attempt to illustrate this in the next three panels of Figure 2, where we plot the fraction of all galaxies with Eddington parameter greater than some fixed value as a function of D$_n$(4000). Dotted, short-dashed, long-dashed and dashed-dotted lines show results for increasing values of the Eddington parameter threshold (see caption).  In the top right panel, we assume that {\em none} of the [OIII] luminosity in galaxies classified as star-forming should be attributed to the AGN. In this case, one sees that the fractions reach a peak value at D$_n$(4000)$ \sim 1.5 -  1.6$. 
They drop steeply at higher values of D$_n$(4000), but they also decline slightly at lower values. 
In the bottom left panel, we make the opposite assumption, i.e. {\em all} of the [OIII] luminosity in galaxies 
classified as star-forming show be attributed to the AGN. These two assumptions obviously bracket 
the full range of possibilities. What is interesting, is that even if we make the extreme 
assumption that all of the [OIII] should be attributed to the AGN, the AGN fraction dependence 
on D$_n$(4000) is remarkably flat below a value of $\sim 1.6$. We will return to this point later. 
The bottom right panel shows an intermediate case in which  
a third of the [OIII] luminosity in the galaxies classified as star-forming
actually comes from the AGN.

In the following analysis, we implement a simple
method for partitioning the total [OIII] luminosity of a galaxy into the component   
from star formation and the component from the AGN, based on the position of the 
 galaxy on the BPT diagram. We assume
 that galaxies situated at  the 
lower ridge  of the star-forming sequence do not contain an AGN.
We then create a set of pure star-forming galaxy
templates at different positions along
the star forming locus by averaging together the emission  line  luminosities  
of galaxies situated along this ridge. Likewise, we create
two different ``pure AGN'' templates by averaging together the emission  line luminosities 
of galaxies situated at the far end of the AGN sequence.
We then create a set of mixing-line "trajectories" by averaging
the emission line luminosities of the star-forming galaxy 
and AGN templates in different proportions.

These trajectories are illustrated in Figure 3. There are 8 curves
shown on the plot, each connecting a point on the star-forming locus to
two possible ``pure'' AGN. The curves are coloured according
to the fraction of the total [OIII] luminosity that originates from 
star formation. Cyan shows the part of the trajectory where the contribution
from star formation is greater than a half, blue  is where it is between
1/3 and 1/2, yellow between 1/4 and 1/3, green between 1/6 and 1/4,
black between 1/6 and 1/9,  red  between 1/9 and 1/16
and magenta is where the SF contribution is less than 1/16th of the total
[OIII] luminosity. For galaxies at the boundary between the star-forming sequence and AGN as
defined by Kauffmann et al. (2003b), the contribution to the [OIII] luminosity by star-formation is predicted to be typically 40 to 50\%. 
By the time, the galaxy has crossed above the "maximum-starburst" curve defined by Kewley et al. (2001), the contribution to [OIII] from star formation has dropped to typically 10-20\%.
\footnote{It is instructive to use the same sets of models to predict the fractional contribution by regions of star formation by other emission-lines in the SDSS spectra. For H$\alpha$, these values are 80-90\% at the Kauffmann et al. (2003b) boundary and 50-60\% at the Kewley et al. (2001) boundary. Values for the other strong emission-lines ([OII]3727, [NII]6584, and [SIII]6717,6731) are similar. The only  line with similar behavior to [OIII]5007 is [OI]6300. However, this line is typically three to ten times weaker than the [OIII] line, and is therefore unable to probe the lowest luminosity AGN.}

To simplify matters, we only use two of the trajectories in shown Figure 3 to estimate the fraction of the total  
[OIII] luminosity from star formation for each galaxy in our sample.
One trajectory is representative of the population of lower ionization
AGN (LINERs), and the other is a better match to higher ionization AGN (Seyferts). 
We calculate the distance to the upper red dot
if the galaxy lies above the dotted red line, or to the lower red dot
of the galaxy lies below this line. We then read off the correction
to the [OIII] luminosity at the same distance along
the trajectory linking the red dot to its corresponding pure AGN.
The correction means that  the fraction of the [OIII] luminosity contributed by star formation 
decreases smoothly with distance from the locus of pure star-forming
galaxies in the BPT diagram. Although this procedure will not be  accurate for
individual galaxies, Figure 3 shows that it ought to work reasonably well
for statistical analyses of the kind presented in this paper.

\section{Eddington Ratio Distribution Functions}

We now turn to an analysis of the distribution function of Eddington parameters. 
We note that our galaxies are drawn from the  $r$-band selected Main spectroscopic sample with $14.5 < r < 17.77$ and in all our work, we are careful to weight by the appropriate $1/V_{max}$ factor to account for any luminosity-dependent selection effects that may be present.   

In Figure 4, we begin by plotting the fraction of black holes with L[OIII]/M$_{BH}$ values above a given value (left panel). We also plot the fraction of the total [OIII] luminosity coming from such black holes (right).
Cyan, blue, green. black, red and magenta curves indicate black holes of increasing mass, from $10^7$ to $10^{8.25} M_{\odot}$ in steps of 0.25 dex in log M$_{BH}$. As discussed above, our adopted bolometric correction implies that an AGN with $\log$ L[OIII]/M$_{BH} \sim 1.7$ is accreting at about the Eddington
rate. As seen in Figures 2 and 3, this value does indeed correspond to the upper
limit of the Eddington parameters of the AGN in our sample.

Figure 4 shows that the duty cycle of moderately powerful AGN activity in low mass black holes is quite high; more than 50\% of all such black holes  have  log L[OIII]/M$_{BH} > -0.2$, corresponding to Eddington ratios of around 1 percent, and 90\% have log L[OIII]/M$_{BH} > -2.2$ . 
In higher mass black holes, the fraction of such moderate luminosity AGN decreases sharply and the population is 
dominated by systems that are accreting at less than 0.1 \% Eddington. Nevertheless, the right-hand
panel of Figure 4 shows that total black hole growth is always dominated by the
most powerful AGN, irrespective of the black hole mass of the galaxy.   
Note that in order to calculate the total black hole growth, we integrate the [OIII] luminosities (corrected for star formation)
of all galaxies with emission lines that are detected at high enough significance to allow them to
be localized on the BPT diagram.

\begin{figure}
\centerline{
\epsfxsize=12cm \epsfbox{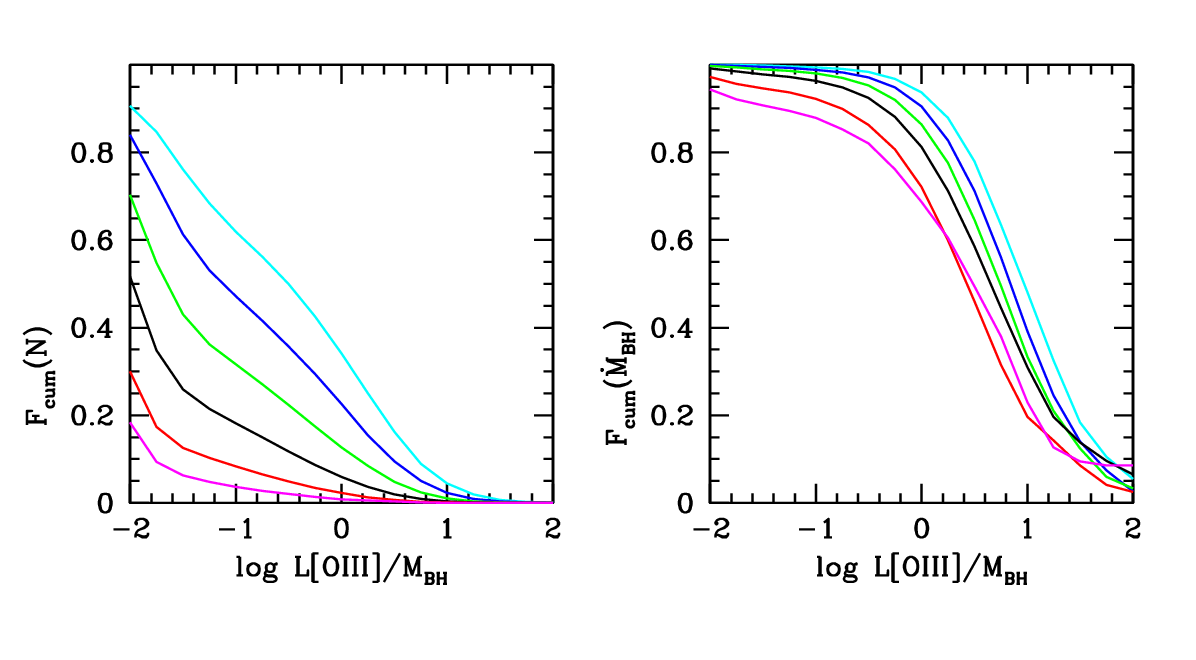}
}
\caption{\label{fig4}
\small
{\it Left:} The fraction of the total number of black holes
with Eddington parameter L[OIII]/M$_{BH}$ greater than a given value. 
{\it Right:} The fraction of the total mass accretion (as traced by
the integrated [OIII] luminosity after correction for star formation) that is occurring in AGN with   
 Eddington parameter L[OIII]/M$_{BH}$ greater than a given value. 
Cyan, blue, green, black, red and magenta curves show results for
black holes with log $M_{BH}$ in the range
(7.0-7.25; 7.25-7.5; 7.5-7.75; 7.75-8.0; 8.0-8.25; 8.25-8.5)}
\end {figure}
\normalsize

We now turn to an analysis of the differential Eddington ratio distribution function, i.e. the fraction of  black holes of a given mass per unit logarithmic interval of   $\lambda$=L[OIII]/M$_{BH}$. Figure 5 shows our derived Eddington parameter distribution functions in 6 different ranges of black hole mass. In this figure, one can clearly see the shift towards lower Eddington parameter as black hole mass increases. The distribution functions have quite complex shapes. For the lower mass black holes, there is a "bump" that is peaked at log [OIII]/M$_{BH}$  of around 0.1 or so. For the higher mass black holes, the bump is barely discernible and the shape is closer to a power-law.

\begin{figure}
\centerline{
\epsfxsize=12cm \epsfbox{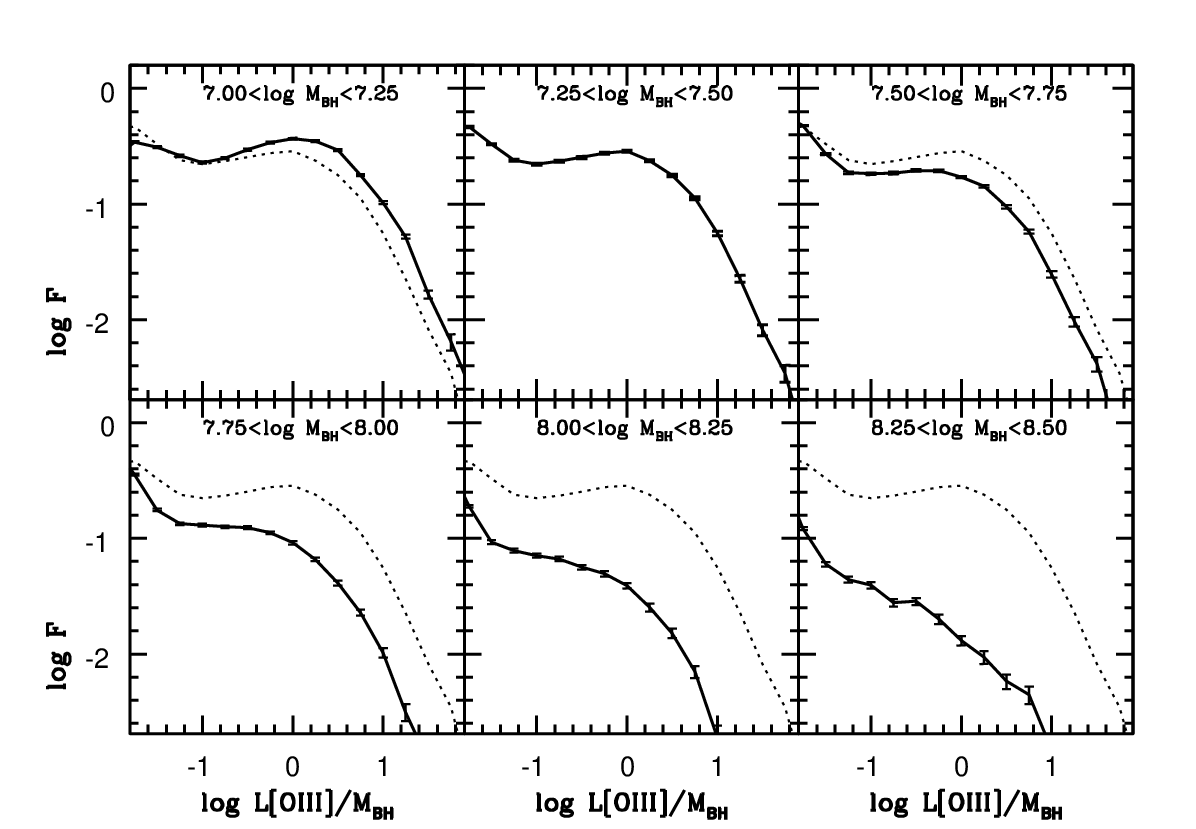}
}
\caption{\label{fig4}
\small
The logarithm of the  fraction of  black holes of a given mass 
per unit logarithmic interval in  $\lambda$=L[OIII]/M$_{BH}$ is plotted as a function of 
log L[OIII]/M$_{BH}$, for six ranges of black hole mass. The dotted line in each panel shows the result
for black holes in the mass range $7.25 < \log M_{BH} < 7.50$.  }
\end {figure}
\normalsize

The origin of this complexity in shape becomes clear when we divide the galaxies in our sample into different bins in 4000 \AA\ break strength. This is shown in Figure 6. The top panel shows Eddington parameter distribution function for all black holes in the range $10^7 -10^8 M_{\odot}$. One can clearly see the bump at high values of log L[OIII]/M$_{BH}$ and the power-law at low values of log L[OIII]/M$_{BH}$. In the two bottom panels, we divide the galaxies into different bins in D$_n$(4000) and the plot the distribution functions for these different subsamples. The left panel shows the results for 5 subsamples with D$_n$(4000)$< 1.5$. It is remarkable that the distribution functions are nearly independent of D$_n$(4000) over this range and are well-characterized by a log-normal function. The right panel shows results for 5 subsamples with D$_n$(4000)$>1.5$. In this regime, the distribution function depends strongly on D$_n$(4000): the fraction of high L[OIII]/M$_{BH}$ objects decreases for larger values of D$_n$(4000) and the shape tends increasingly towards a power-law. 

\begin{figure}
\centerline{
\epsfxsize=13cm \epsfbox{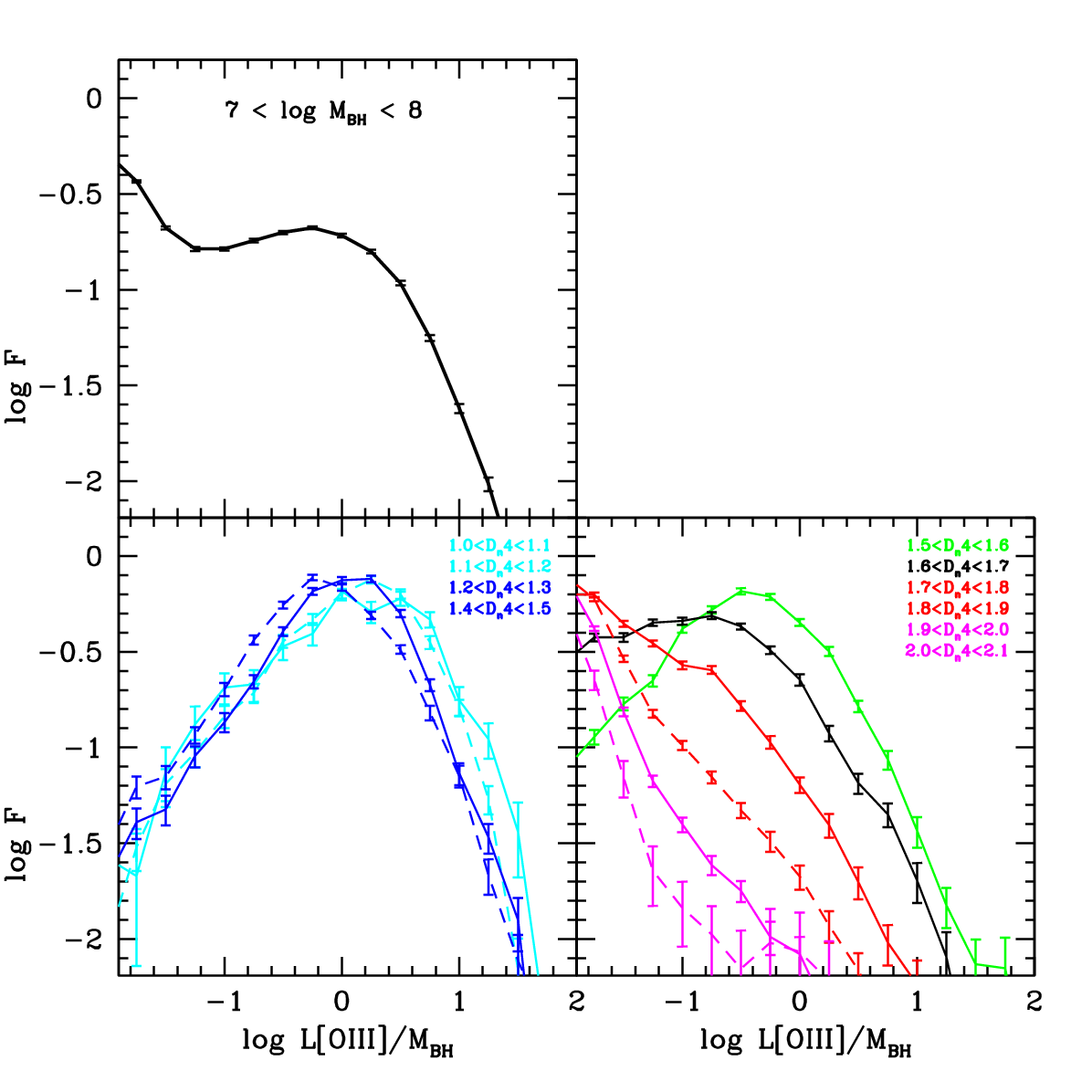}
}
\caption{\label{fig4}
\small
{\it Top:} The logarithm of the  fraction of  black holes per unit 
logarithmic interval in  $\lambda$=L[OIII]/M$_{BH}$, and  with masses
in the range $10^7 -10^8 M_{\odot}$ and 
is plotted as a function of 
log L[OIII]/M$_{BH}$. {\it Bottom:} The accretion rate distributions
of  subsamples split into different ranges in 4000 \AA\ break
strength as indicated.  }
\end {figure}
\normalsize

In Figure 7, we explore how these results depend on the mass of the black hole.
We have chosen two representative ranges in 4000 \AA\ break strength: $1.2<$D$_n$(4000)$<1.4$, corresponding to 
"young" galaxies, and $1.7<$D$_n$(4000)$<1.9$, corresponding to "old" galaxies. We plot the distribution 
functions of our Eddington ratio parameter for different ranges of black hole mass. We see that the 
log-normal distribution is not only independent of the precise value of 4000 \AA\ break strength for
 galaxies with D$_n$(4000)$< 1.5$ (Figure 6), it is also independent of black hole mass. 
The best fit log-normal function is centered at a value of log L[OIII]/M$_{BH}$=0.1
and has a dispersion of 0.4 dex.  As shown 
in the right panel, however, the slope of the power-law is 
roughly constant ($\sim -0.7$) for different mass black holes, but there is a dependence in the amplitude --  
Eddington ratios are apparently lower for galaxies 
with more massive black holes.

\begin{figure}
\centerline{
\epsfxsize=12cm \epsfbox{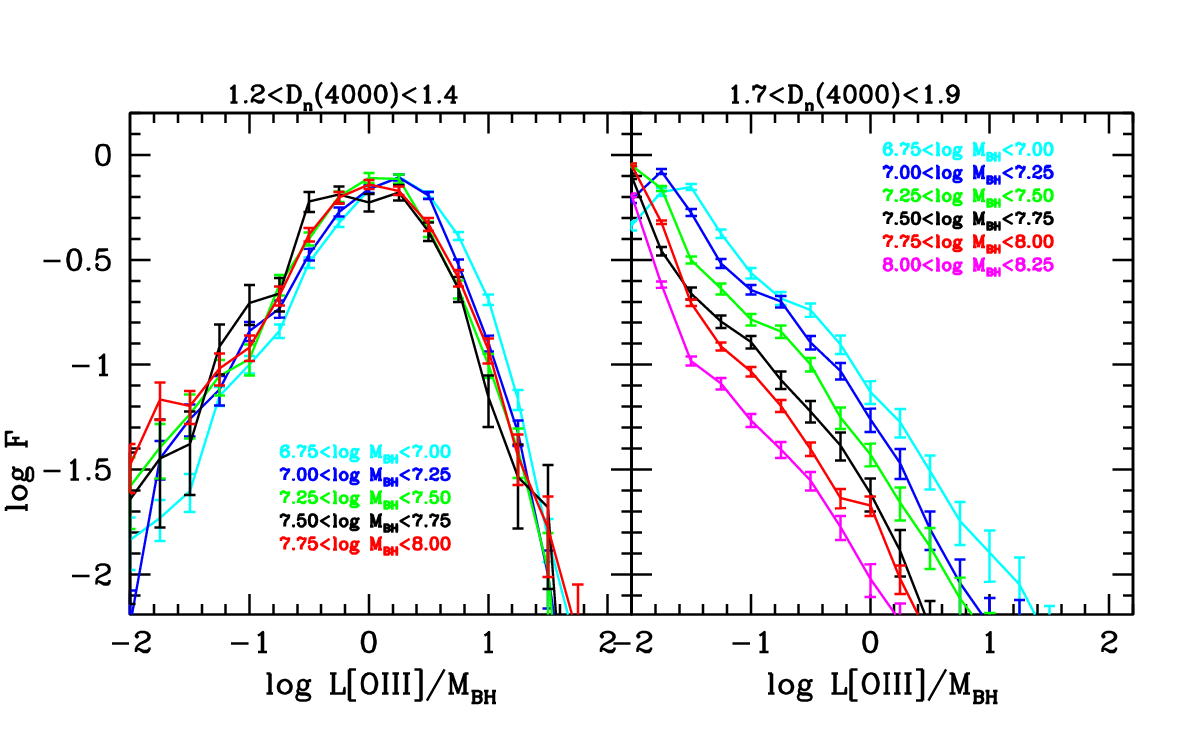}
}
\caption{\label{fig4}
\small
The logarithm of the  fraction of  black holes per unit
logarithmic interval in  $\lambda$=L[OIII]/M$_{BH}$,  with masses
in the range $10^7 -10^8 M_{\odot}$ a
is plotted as a function of 
log L[OIII]/M$_{BH}$ for different ranges of black hole mass. The left panel shows
results 
for young galaxies with $1.2<$D$_n$(4000)$<1.4$  and
the right panel shows results  for old galaxies
with $1.7<$D$_n$(4000)$<1.9$ (right). The Eddington parameter distribution in the left panel
is well-fit with a log-normal function centred at log L[OIII]/M$_{BH}$=0.1
and with a dispersion of $\sim 0.4$ dex. In the right panel, the distribution can be characterized
as a power-law with slope $\sim -0.6 - -0.8$. }
 
\end {figure}
\normalsize

Although it is clear from Figure 7 that the shape of the accretion rate distribution functions are quite distinct in galaxies with "young" and "old" stellar populations, it is important to note that {\em the accretion rate
distribution is very broad in both populations}. Indeed, in galaxies with
black hole masses less than a few $\times 10^7 M_{\odot}$, the fraction of systems that are accreting near Eddington differs rather little  between the two populations. The main distinguishing characteristic is that for the young population, the distribution peaks at a characteristic value of about 2.5 percent Eddington \footnote {We note that
the precise position of the peak is  sensitive to how we choose to correct the
[OIII] luminosities of the galaxies in our sample for star formation.
For example, if we simply assign a third
of the [OIII] luminosity of all galaxies lying below the Kauffmann et al (2003b)
demarcation curve to the AGN, and 100\%  of the [OIII] luminosity
from galaxies above this curve to  the AGN, we obtain a broader 
distribution peaked at 0.5 percent Eddington. However, the 
bell-like shape of the distribution function and our conclusion that it 
peaks at values much less than Eddington is robust. It holds even if
if we assign  100\% of the [OIII] luminosity in all
galaxies to the AGN. }    
 , but this peak is not present for galaxies with old stellar populations. This means that the likelihood for the black holes in young galaxies  to be accreting at rates very close to Eddington is small , but so
is the probability for such black holes to be accreting at very low rates or not
to be accreting at all.  On the other hand, black holes in galaxies with 
old stellar populations spend most of their time inactive or  accreting at very low rates, but accretion near the Eddington luminosity is still possible under certain (rare) conditions. This point is again apparent in Figure 8, where we 
show plots of the cumulative fraction of the total mass accretion onto black holes that is occurring above a given value of the Eddington parameter,
in six different ranges of 4000 \AA\ break strength. The different coloured lines show results for different ranges of black hole mass. We see that over most of the range in black hole mass and stellar age, the growth of black holes is dominated by the most actively growing systems (50\% of the growth is occurring in systems accreting at greater than 5 to 10\% of the Eddington limit). However, for the biggest black holes in the oldest bulges, this median value drops to only 0.1 to 1\% Eddington.

Finally, Figure 9 shows the fraction of the total mass accretion occurring in the
"log-normal" (defined as D$_n$(4000)$< 1.7$) versus the "power-law"
(defined as D$_n$(4000)$>1.7$) regimes as a function of black hole mass.
For low mass black holes, $\sim$ 80\% of the accretion is occuring in the
log-normal mode. This fraction decreases for more massive black holes -- 
the two modes of black hole growth become  roughly equal for black holes
of $\sim 10^8 M_{\odot}$. At larger black hole masses, the "power-law"
mode dominates. Integrating over black holes of all masses, we find
that 76\% of the total mass accretion in the local Universe is occuring
in the log-normal regime at the present day.  Thus, both modes of black hole growth are significant.

\begin{figure}
\centerline{
\epsfxsize=15cm \epsfbox{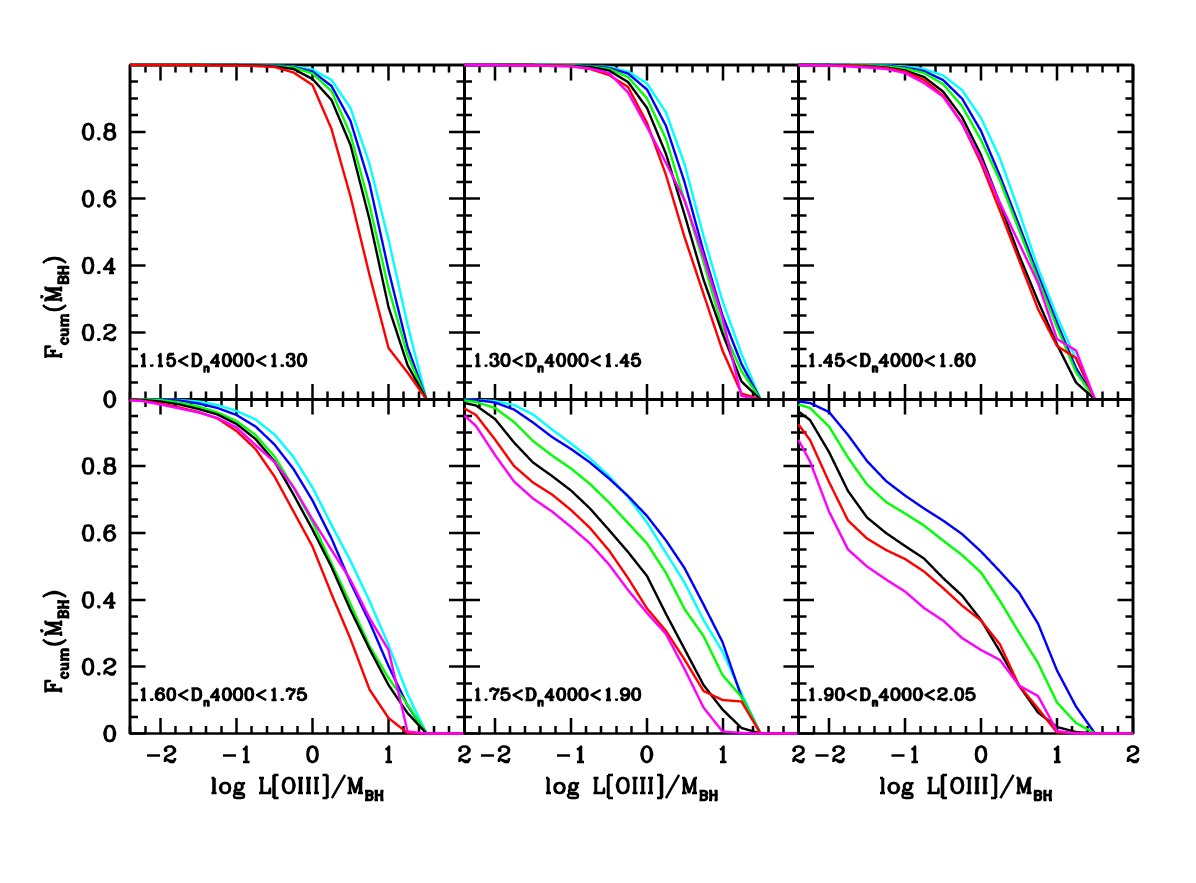}
}
\caption{\label{fig4}
\small
The fraction of the total mass accretion (as traced by
the integrated [OIII] luminosity) that is occurring in AGN with   
Eddington parameter L[OIII]/M$_{BH}$ greater than a given value. 
Different panels are for subsamples in different ranges of 4000
\AA\ break strength.
Cyan, blue, green, black, red and magenta curves show results for
black holes with log $M_{BH}$ in the range
(7.0-7.25; 7.25-7.5; 7.5-7.75; 7.75-8.0; 8.0-8.25; 8.25-8.5)}
\end {figure}
\normalsize

\begin{figure}
\centerline{
\epsfxsize=8cm \epsfbox{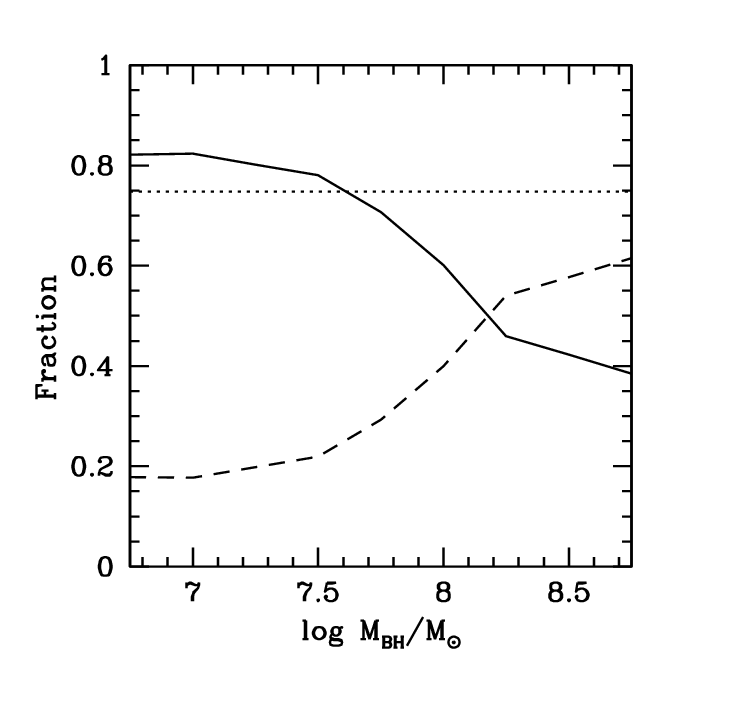}
}
\caption{\label{fig4}
\small
The fraction of the total mass accretion (as traced by
the integrated [OIII] luminosity) that is occurring in  the       
"log-normal" mode (solid line) and the "power-law" mode
(dashed line) as a function of black hole mass. The dotted
line shows the fraction of the total mass accretion
onto all black hole that is occurring the 
log-normal mode.}
\end {figure}
\normalsize

\section {Towards a Physical Understanding of these Results}

Our basic result is that we have uncovered two regimes of black hole growth in
the local Universe. One regime is associated with galaxies with D$_n$(4000) $< 1.5$ (galaxies with significant recent/on-going star formation) and is characterized by a log-normal distribution function of accretion rates that is peaked at value of L[OIII]/M$_{BH}$ corresponding to a few   
percent of the Eddington rate. The other regime is associated with galaxies with D$_n$(4000) $> 1.8$ (little or no star formation) and is characterized by a power-law-like distribution function of accretion rates, extending down to the detection threshold of the [OIII] line in our SDSS spectra.     

\subsection {The log-normal regime}

Our most striking conclusion about the log-normal regime is that the accretion rate distribution function 
remains largely invariant if we adopt different cuts on galaxy properties. It is independent of the velocity dispersion of the stars in the bulge and by extension, it is independent of the mass of the black hole. It is also
at most weakly dependent on the fraction of young 
stars measured within the SDSS fibre aperture, which implies that the accretion onto the black hole is insensitive to the properties of the interstellar medium
as measured inside radii of $\sim 1$ kpc. 

Thus in the log-normal regime,  accretion onto the black hole is apparently not limited by the supply of gas, but by feedback processes that are intrinsic to the black hole itself, such as winds from the accretion disk (Begelman, McKee \& Shields 1983; Murray et al 1995) 
or small-scale radio jets (Ulvestad \& Wilson 1989; Kellerman et al 1998). One could imagine a situation where the black hole flares to a luminosity near Eddington when it first accretes some gas and this is followed by a period of latency, because feedback processes disrupt the gas supply in the vicinity of the black hole. Indeed, the location of the peak of the distribution at a few percent of Eddington may
reflect an accretion rate threshold where the feedback starts to become important.  
It has been pointed out by many authors that radiative feedback effects are predicted  to reduce steady state accretion onto 
the black hole to $\sim 1$\%  Eddington (e.g. Ostriker et al. 1976, Proga et al 2008, Milosavljevic et al 2008).
This main difference between our picture and that envisgaed by  Hopkins \& Hernquist (2008) 
is that the feedback
is  not strong enough to disrupt and expel the entire interstellar medium of the galaxy. If this was the case, one would not obtain a log-normal distribution of accretion rates, with a decrease in the fraction of objects accreting at low luminosities  relative to those accreting at intermediate luminosities. In low redshift AGN the bulk of the gas in the bulge remains intact, and after a period of time that is short compared to the age of the Universe, gas will again begin to accrete onto the black hole.

These results help us understand why no clear connection has been established between the presence of central bars or dust lanes and the fueling of AGN in local galaxies (e.g. Martini et al. 2003). If the effect of feedback from the black hole is limited only to the gas very near the black hole, and if the AGN lifetime is relatively brief, then we would not expect to see systematic differences in the kpc-scale properties of young bulges with and without a strong AGN. These results also help us understand why there is no connection between mergers/interactions and black hole accretion rate in AGN in the local Universe. Galaxy-galaxy interactions drive gas from the outer disk into the central kiloparsec of the galaxy, where  it can fuel a starburst. Indeed, we do see strong signatures of this process in the SDSS data (Li et al 2008; Reichard et al 2008). However, this increase in gas supply has no significant impact on the accretion onto the central black hole for reasons that we have discussed above.

\subsection {The power-law regime}
At first sight, the power-law regime appears more complicated. Figures 6 and 7 show that the amplitude of the power law appears to depend on both black hole mass and on D$_n$(4000). However, in Figure 10, we show that the picture simplifies considerably if we plot distribution functions of the 
quantity L[OIII]/M(bulge), where M(bulge)  is the dynamical mass of the  galaxy bulge, which we estimate
using  the formula M(bulge)= $(1.65\sigma)^2 R_{50}/G$, where $R_{50}$ is the half-light radius of the galaxy (see Padmanabhan et al 2004).
Figure 10 shows that the slope of the power-law ($\sim$ -0.8) is similar to before, 
but now the distributions are no longer dependent on black hole mass 
and only scale with D$_n$(4000).

This implies that in the power-law regime, the total accretion onto the black hole is simply proportional to the stellar mass in the bulge, with a constant of proportionality that scales with the mean stellar age of the stars in the bulge.
The simplest interpretation of this result is that black holes in this regime 
are being fueled by stellar mass loss.

Norman \& Scoville (1988) discussed a model in which stellar mass 
loss from an evolving nuclear starburst fueled the growth of a central 
super-massive black hole. They found that the evolution of the 
stellar mass-loss over the range from 100 Myr to 10 Gyr after the 
burst could be parameterized as a power-law in time with a slope 
in the range -1.1 to -1.5 (leading to the same scaling for the 
luminosity of the accreting black hole). Following 
Hogg \& Phinney (1997), the resulting slope of the luminosity 
function $dN/dlnL$ for this population would be in the range 
-0.7 to -0.9. This is in excellent agreement with our measured 
slope for power-law distribution of the Eddington ratio.

We can further test our model by using 
standard population synthesis models (Bruzual \& Charlot 2003; Maraston 2005) to predict how
mass loss rates ought to scale as a function of D$_n$(4000), 
and to investigate whether this agrees with the change in the 
average value of L[OIII]/$M_*$ as a function of D$_n$(4000) 
for the AGN in our sample. This is shown in Figure 11. The  dots 
show  the predicted mass loss rate (in units of the fraction of the stellar mass returned per Gigayear) as a function of D$_n$(4000) for galaxies with different star formation histories. We parametrize these histories using a range of formation times, exponential decline time scales
and burst probabilities as described in Kauffmann et al (2003a). Black dots show results using 
mass loss rates taken from Bruzual \& Charlot (2003), while magenta dots are for
Maraston (2005). A Kroupa (2001) IMF is assumed in both cases. 
The coloured curves show how the average value of L[OIII]/M(bulge) 
changes  as a function of D$_n$(4000) for galaxies with black holes in different mass ranges. 
As can be seen, the predicted  mass loss rates decrease by a factor of $\sim 3$ as D$_n$(4000) 
goes from a value of 1.6 to a value of 2.1. The corresponding drop in L[OIII]/M$_*$ is somewhat larger, 
but taking into account the uncertainties in the IMF in the dense inner regions of galaxy bulges (Maness et al 2007), 
the fact that the two quantities scale to within a factor of two is encouraging.

\begin{figure}
\centerline{
\epsfxsize=13cm \epsfbox{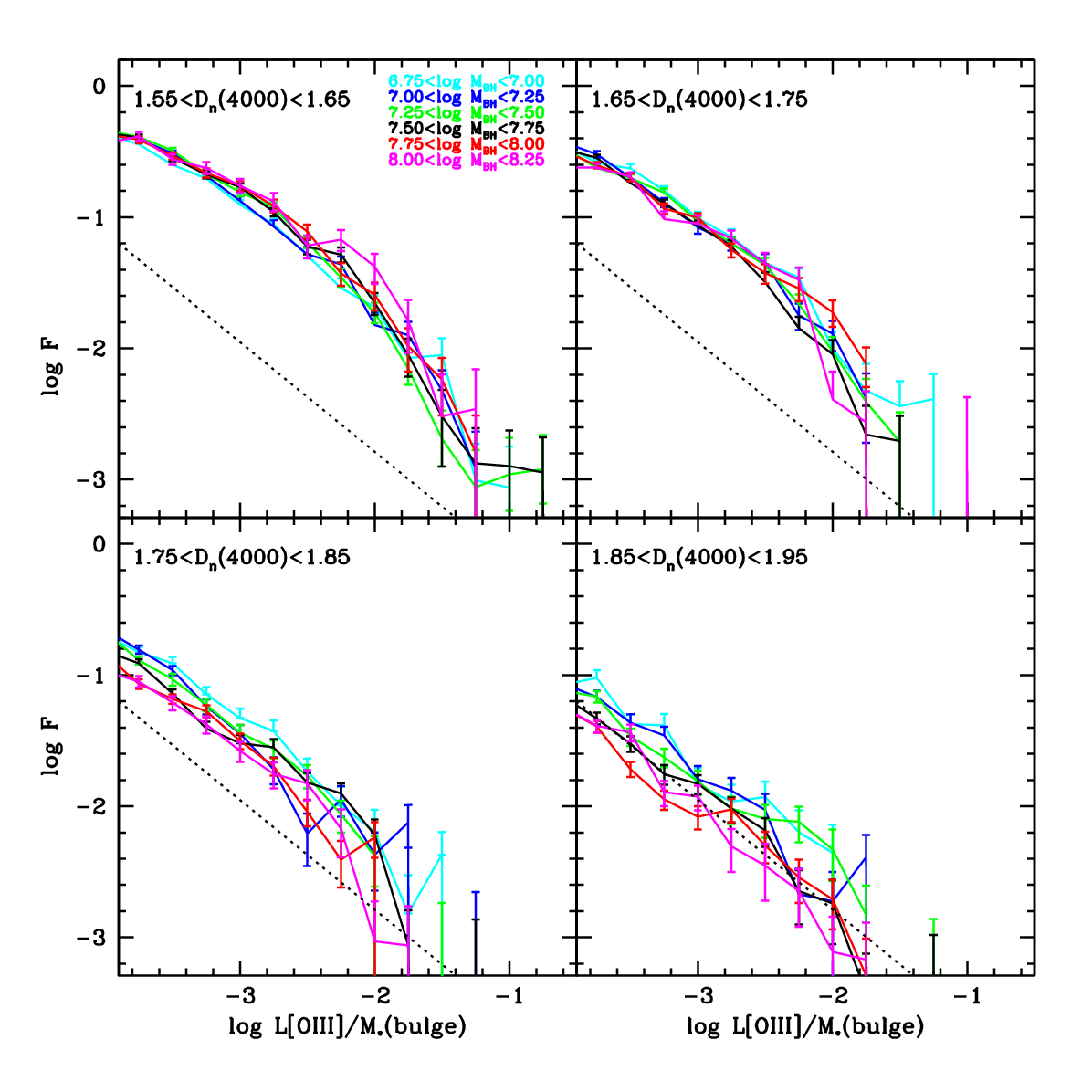}
}
\caption{\label{fig4}
\small
The distribution function of the ratio of the [OIII] line luminosity to bulge dynamical mass 
for galaxies split by black hole mass and by 4000 \AA\ break strength.
We plot the fraction of black holes per unit logarithmic interval
in log L[OIII]/M(bulge).
The 4 different panels show results for four different
ranges in 4000 \AA\ break strength,
Cyan, blue, green, black, red and magenta curves show results for
black holes with log $M_{BH}$ in the range
(7.0-7.25; 7.25-7.5; 7.5-7.75; 7.75-8.0; 8.0-8.25; 8.25-8.5). The dotted black line shows a power-law with a slope of -0.8 adjusted to pass through the data in the lower right panel.}
\end {figure}
\normalsize

\begin{figure}
\centerline{
\epsfxsize=11cm \epsfbox{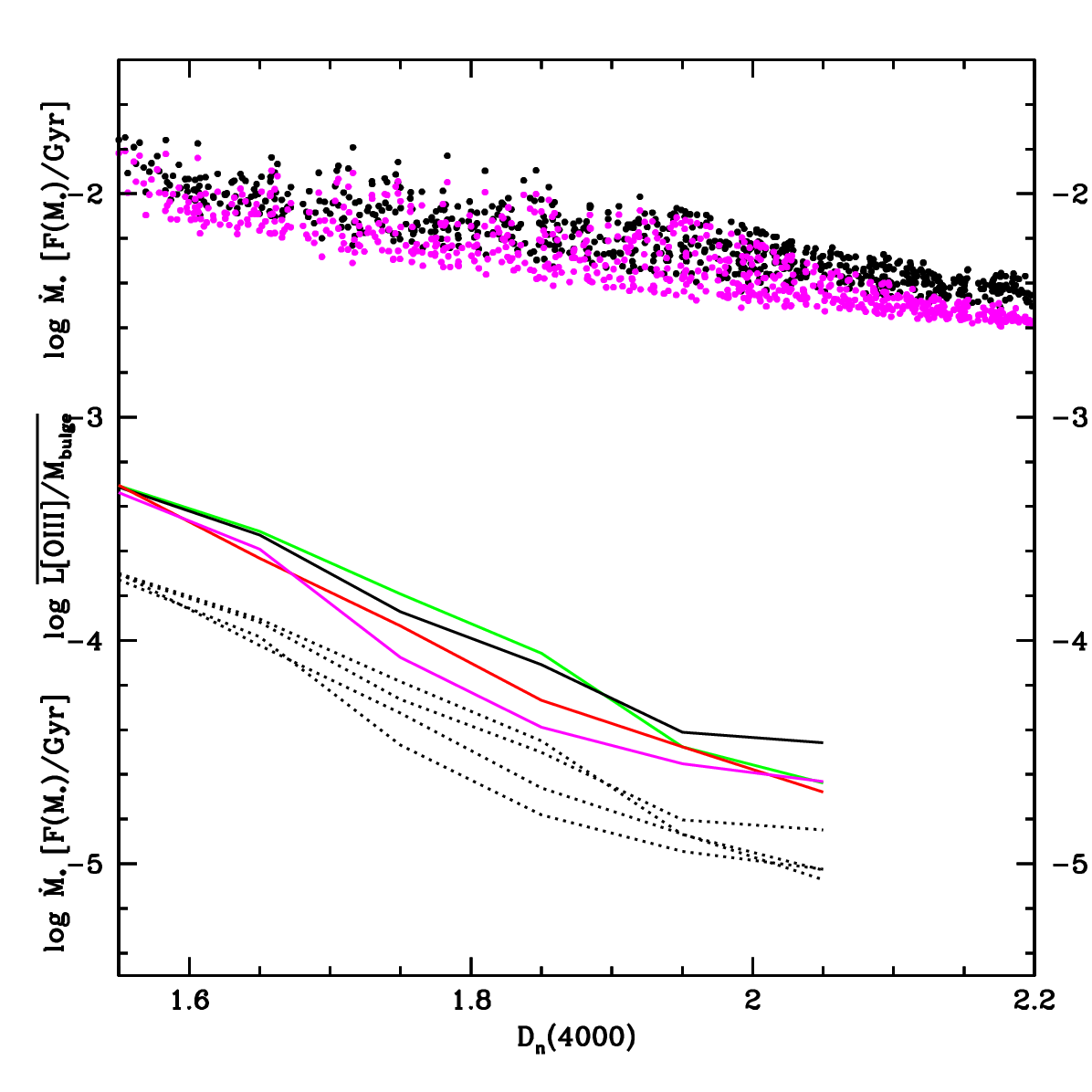}
}
\caption{\label{fig4}
\small
The black dots show the mass loss rate per Gyr normalized to the stellar mass predicted by BC03 models (black) and Maraston (2005) models
(magenta) as a function of D$_n$(4000)
for a variety of exponentially declining star formation histories.  
The coloured curves show the mean  value of L[OIII]/$M(bulge)$ (in units of solar luminosities
per solar mass).
as a function of D$_n$(4000) for galaxies with older stellar populations.
Green, black, red and magenta curves show results for
black holes with log $M_{BH}$ in the range
(7.25-7.5; 7.5-7.75; 7.75-8.0; 8.0-8.25).
The dotted black curves show the implied mass accretion rate 
onto the black hole per Gyr (again normalized to the stellar mass), if we assume a bolometric correction
factor $L_{bol}/L_{[OIII]} \sim 600$ and a radiative efficiency of 10\%. The ratio of the normalized accretion rates for these dotted curves to those of the models (black and magenta points) indicates the fraction of the stellar mass loss that would need to be accreted by the black hole. The implied ratios are $\sim$ 0.3 to 1\%.}
\end {figure}
\normalsize

We have also tried scaling the [OIII] luminosities by the stellar mass of the whole 
galaxy and we find L[OIII]/M$_*$ distributions functions that are still independent of black hole mass, 
but with somewhat  more scatter. It is important to recall that our analysis always deals with average accretion rates, and not accretion rates for individual AGN. 
 Because we are considering accretion rates averaged over a large population of objects, it is not unreasonable to suppose that if the main source of fuel is mass loss from evolved stars, the average accretion rate will scale with the stellar mass of the galaxy bulge. On the other hand, if one were able to  
"zoom in" very close to the  active  nucleus, then the {\em instantaneous} accretion rate onto the black hole and the mass and mean age of the  stars in the {\em immediate vicinity} of the nucleus may  correlate with much less scatter. 
Recently, Davies et al (2007) have analyzed star formation in the nuclei of
nine Seyfert galaxies at spatial resolutions corresponding to length scales of
around 10 pc. They find evidence for young stars in the near vicinity of all 
the nuclei, but in many of the systems, star formation has already 
terminated. These results suggest a scenario in which nuclear star 
formation occurs in multiple short bursts, as gas (perhaps originating 
from the material shed by stars further out in the bulge) flows 
into the central regions of the galaxy. Interestingly, the luminosity of the AGN appears to be {\em positively}  correlated with the age of the starburst, 
peaking at ages of 100-300 Myr. It is then argued that this may 
imply that winds from AGB stars, which have low velocities, 
may supply the gas which actually flows onto the  central black hole. 
This time delay agrees with the prediction of Norman \& Scoville (1988).

Finally, we use our fiducial bolometric correction factor $L_{bol}$/L[OIII]$ \sim 600$, plus an assumed radiative efficiency of 10\%, to transform the average L[OIII]/M$_*$ values into estimates of the average black hole accretion rate. This is shown as the set of dotted curves at the bottom of the plot
(again, per Gyr normalized to the stellar mass of the bulge). The ratio of this accretion rate to the mass-loss rate predicted by the models yields the fraction of the mass-loss from the stellar bulge that must be accreted by the black hole to produce the observed line radiation. Figure 11 then implies that this fraction is between 0.3-1\%. This is in good agreement with  Ciotti \& Ostriker (2007), who model how AGN activity is triggered by recycled gas from evolved stars. These models predict that half the gas from dying stars is ejected as galactic winds, half is consumed in central starbursts and less than 1\% is accreted onto the central black hole. The Norman \& Scoville (1988) model assumed much higher accretion efficiency, although their source of the mass-loss was a compact (10 pc-scale) nuclear star cluster, and they did not consider feedback from the black-hole.

\section {Final Thoughts \& Conclusions}

As we have shown,  the changeover from AGN in the log-normal regime to AGN in the power-law regime 
starts to occur at a 4000 \AA\ break strength of $\sim$ 1.5 and is complete by values 
above $\sim$1.8. We can use the calibration of D$_n$(4000) in terms of the star formation 
rate per unit mass in Brinchmann et al. (2004) to examine the physical significance of these values. 
We use Figure 11 in Brinchmann et al., but consider the inverse quantity: M$_*$/SFR 
(i.e. the characteristic timescale for growing the bulge due to star formation). 
This figure then shows that over the range in D$_n$(4000) spanned by the log-normal regime, 
this growth time increases from about 5 Gyr (for D$_n$(4000) = 1.2) to about 20 Gyr 
(for D$_n$(4000) = 1.5). In other words, the log-normal regime corresponds to galaxy bulges 
that are growing significantly today (M$_*$/SFR $\sim$ the Hubble time). This figure also shows 
that in the pure power-law regime (D$_n$(4000) $>$ 1.8), the growth time is about two 
orders-of-magnitude longer than the Hubble time (i.e. these are dead systems). Finally, 
the intermediate regime of D$_n$(4000) = 1.5 to 1.8 where AGN transition from the 
pure log-normal to pure power-law regime corresponds to a similar transition 
between systems with short and very long growth times (from $\sim$ 20 to 1000 Gyr). 
In simple terms, the behavior of the distribution of the Eddington parameter implies
 that the log-normal and power-law regimes correspond almost exactly to the populations 
of "living" and "dead" galaxy bulges.

In conclusion, therefore, optical emission lines trace  two distinct modes of black hole growth in the
present-day universe. 
We have shown that these two modes are 
strongly linked to the formation history of the stellar population of the bulges that host  
the black holes. These links provide important clues as to how  black holes are  fueled 
and how this fueling may regulate their  further growth.

One mode is associated with galaxy bulges that are undergoing significant star-formation. 
More quantitatively, these are bulges in which the present stellar mass can be 
formed at the current star-formation rate on a timescale of-order the Hubble time. 
Thus, these are systems in which the build up of the stellar population has taken place 
gradually over the history of the universe.  For these black holes, the 
distribution of the Eddington-ratio ($L/L_{Edd}$) takes a log-normal form that 
is universal with respect to both the mass of the black hole and the current star formation rate.
This regime dominates the growth of black holes with masses less than $\sim10^8$ M$_{\odot}$. 
As shown by Heckman et al. (2004),  the mass-doubling time for this population of black 
holes is of-order a Hubble time. 

The other mode is associated with galaxy bulges with little or no on-going 
star-formation. In these systems, the formation of the stars in the bulge occurred long ago. 
For this population of black holes, the distribution of the Eddington-ratio is
a power-law with a universal slope, but with a normalization that depends on the age of 
the stellar population. We have shown that the slope and age-dependence of the normalization 
are predicted by a simple model in which the black hole accretes on-average 
$\sim$ 0.3 to 1\% of the mass lost by evolved bulge stars. This regime dominates the growth 
of the more massive black holes ($>10^8$ M$_{\odot}$), the population with an 
insignificant present-day growth rate (Heckman et al. 2004). 

As has been discussed by many authors, the galaxy population is the local Universe is also strongly bimodal 
in nature, with galaxies separating into two clearly defined "peaks", both in terms of their 
structure and their stellar populations (Strateva et al 2002; Kauffmann et al 2003c; Baldry et al 2004). 
This has led to considerable speculation as to  
the physical mechanisms that are responsible for creating the two distinct populations.
One idea is that feedback from the AGN may be powerful enough to expel much of the interstellar medium 
of the galaxy, causing star formation to shut down over short time scales and the galaxy to 
transit from the blue to the red population. 
We have argued that the log-normal Eddington ratio distribution functions for galaxies with young bulges are 
inconsistent with catastrophic feedback from the black hole 
that operates on the scale of the whole galaxy (or even of the bulge). 
Instead, it is likely that the log-normal distribution of the Eddington ratio is 
set by a long-running competition between the sporadic supply of cold gas from the 
bulge to the black hole and subsequent feedback that operates locally 
(speculatively, within the sphere of influence of the black hole  $r <$ GM$_{BH}/\sigma_{bulge}^2$).
Once the gas runs out, black hole growth appears to be regulated by the 
rate at which evolved stars lose their mass.
The picture of black hole growth that appears to emerge from our results 
is thus very much a "passive" one.

One question one might ask is whether the low redshift AGN studied in this
paper are fundamentally different in nature to quasars at higher redshifts.
Kollmeier et al (2006) concluded that in higher redshift quasars,
the distribution of estimated Eddington ratios is also well described 
as log-normal, but with  a peak at much higher
luminosities ($L_{\rm bol}/L_{\rm Edd}=1/4$).
As pointed out in a number of papers
(e.g. Netzer et al 2007, Gavignaud et al 2008, Hopkins et al 2009), 
selection effects imply only objects with uniformly high accretion rates will be optically identified as quasars.
Only when we have complete samples of {\em galaxies} selected by
bulge or black hole mass 
where the energetic output of the central black holes 
can be characterized down to reasonably low levels,   
will we be able to make a fair comparison between low and high redshift samples.
  
For the moment, we believe that the origin of the bimodal nature of the 
galaxy population still remains as a major unsolved problem. 
Given the continued cosmological infall of gas, why is the only fuel available to  black holes in massive galaxies 
material shed by old stars?  At least at low-redshifts, it is likely that 
radio AGN play an active role in suppressing the cooling of gas in the dark matter halo, 
keeping such systems in the power-law "slow starvation" regime (Best et al. 2005, 2007).
It remains to be seen whether the same picture will still hold up at higher redshifts.

\vspace {0.5cm}

\noindent
\large
{\bf Acknowledgments}\\
\normalsize

We thank Philip Hopkins, Reinhard Genzel, Colin Norman, 
Jerry Ostriker,  Simon White and Francesco Shankar
for useful conversations that inspired this work.

Funding for the SDSS and SDSS-II has been provided by the Alfred
P. Sloan Foundation, the Participating Institutions, the National
Science Foundation, the U.S. Department of Energy, the National
Aeronautics and Space Administration, the Japanese Monbukagakusho, the
Max Planck Society, and the Higher Education Funding Council for
England. The SDSS Web Site is http://www.sdss.org/.

The SDSS is managed by the Astrophysical Research Consortium for the
Participating Institutions. The Participating Institutions are the
American Museum of Natural History, Astrophysical Institute Potsdam,
University of Basel, University of Cambridge, Case Western Reserve
University, University of Chicago, Drexel University, Fermilab, the
Institute for Advanced Study, the Japan Participation Group, Johns
Hopkins University, the Joint Institute for Nuclear Astrophysics, the
Kavli Institute for Particle Astrophysics and Cosmology, the Korean
Scientist Group, the Chinese Academy of Sciences (LAMOST), Los Alamos
National Laboratory, the Max-Planck-Institute for Astronomy (MPIA),
the Max-Planck-Institute for Astrophysics (MPA), New Mexico State
University, Ohio State University, University of Pittsburgh,
University of Portsmouth, Princeton University, the United States
Naval Observatory, and the University of Washington.

\pagebreak
\Large
\begin {center} {\bf References} \\
\end {center}
\normalsize
\parindent -7mm
\parskip 3mm

Adelman-McCarthy J.~K., et al., 2006, 
ApJS, 162, 38 

Baldry I.~K., Glazebrook K., Brinkmann J., Ivezi{\'c} {\v Z}., Lupton 
R.~H., Nichol R.~C., Szalay A.~S., 2004, ApJ, 600, 681 

Baldwin J.~A., Phillips M.~M., Terlevich R., 1981, PASP, 93, 5 

Balogh, M.L., Morris, S.L., Yee, H.K.C., Carlberg, R.G., Ellingson, E., 1999,
ApJ, 527, 54

Barnes J.~E., Hernquist L.~E., 1991, ApJ, 370, L65 

Barnes J.~E., Hernquist L., 1996, ApJ, 471, 115 

Begelman M.~C., McKee C.~F., Shields G.~A., 1983, ApJ, 271, 70 

Best P.~N., Kauffmann G., Heckman T.~M., Ivezi{\'c} {\v Z}., 2005, MNRAS, 
362, 9 (B05b)

Best P.~N., von der Linden A., Kauffmann G., Heckman T.~M., Kaiser C.~R., 
2007, MNRAS, 379, 894 

Brinchmann J., Charlot S., White S.~D.~M., 
Tremonti C., Kauffmann G., Heckman T., Brinkmann J., 2004, MNRAS, 351, 1151 

Bruzual G., Charlot S., 2003, MNRAS, 344, 1000 

Cid Fernandes R., Gu Q., Melnick J., 
Terlevich E., Terlevich R., Kunth D., Rodrigues Lacerda R., Joguet B., 
2004, MNRAS, 355, 273 

Ciotti L., Ostriker J.~P., 1997, ApJ, 487, L105 

Ciotti L., Ostriker J.~P., 2007, ApJ, 665, 1038 

Croton D.~J., et al., 2006, MNRAS, 365, 11 

Davies R.~I., Mueller S{\'a}nchez F., Genzel R., Tacconi L.~J., Hicks 
E.~K.~S., Friedrich S., Sternberg A., 2007, ApJ, 671, 1388 

Di Matteo T., Springel V., Hernquist 
L., 2005, Nature, 433, 604 

Garc{\'{\i}}a-Burillo S., Combes F., Schinnerer E., Boone F., Hunt L.~K., 2005, A\&A, 441, 1011 

Gavignaud, I. et al, 2008, A\&A, 492, 637

Granato G.~L., De Zotti G., Silva L., 
Bressan A., Danese L., 2004, ApJ, 600, 580 

Hao L., et al., 2005, AJ, 129, 1783 

Heckman T.~M., Kauffmann G., Brinchmann 
J., Charlot S., Tremonti C., White S.~D.~M., 2004, ApJ, 613, 109 

Ho L.~C., Filippenko A.~V., Sargent W.~L.~W., 1997, ApJS, 112, 315 

Ho L.~C., 2008, ARA\&A, 46, 475 

Hogg, D.~W., Phinney, E.~S., 1997, ApJ, 488, L95

Hopkins P. F., Hernquist L., Cox, T.J., Di Matteo T., Martini P.,
Robertson, B., Springel, V., 2005a, ApJ, 630, 705

Hopkins P.~F., Hernquist L., Martini P., 
Cox T.~J., Robertson B., Di Matteo T., Springel V., 2005b, ApJ, 625, L71 

Hopkins P.~F., Hernquist L., 2008, arXiv, arXiv:0809.3789

Hopkins, P.F., Hickox, R., Quataert, E., Hernquist, L. 2009, arXiv:0901.2936

Kauffmann G., Haehnelt M., 2000, MNRAS, 311, 576 

Kauffmann G., et al., 2003a, MNRAS, 341, 33 

Kauffmann G., et al., 2003b, MNRAS, 346, 1055 

Kauffmann G., et al., 2003c, MNRAS, 341, 33 

Kellermann K.~I., Vermeulen R.~C., Zensus 
J.~A., Cohen M.~H., 1998, AJ, 115, 1295 

Kewley L.~J., Dopita M.~A., Sutherland R.~S., Heisler C.~A., Trevena J., 
2001, ApJ, 556, 121 

Kewley L.~J., Geller M.~J., Jansen R.~A., Dopita M.~A., 2002, AJ, 124, 3135 

Kewley L.~J., Groves B., Kauffmann G., Heckman T., 2006, MNRAS, 372, 961 

Kollmeier J.~A., et al., 2006, ApJ, 648, 128

Kroupa, P., 2001, MNRAS, 322, 231

Li C., Kauffmann G., Wang L., White S.~D.~M., Heckman T.~M., Jing Y.~P., 2006, 
MNRAS, 373, 457 

Li C., Kauffmann G., Heckman T.~M., White S.~D.~M., Jing Y.~P., 2008, MNRAS, 
385, 1915 

Maness H., et al., 2007, ApJ, 669, 1024 

Maoz, D., 2007, MNRAS, 377, 1696

Maraston C., 2005, MNRAS, 362, 799 

Martini P., Weinberg D.~H., 2001, ApJ, 547, 12 

Martini P., Regan M.~W., Mulchaey J.~S., 
Pogge R.~W., 2003, ApJ, 589, 774 

Mathews W.~G., Baker J.~C., 1971, ApJ, 170, 241 

Mihos J.~C., Hernquist L., 1996, ApJ, 464, 641

Milosavljevic, M., Couch, S.M., Bromm, V., 2008, eprint arXiv:0812.2516

Murray N., Chiang J., Grossman S.~A., Voit G.~M., 1995, ApJ, 451, 498

Netzer, H., Lira, P., Trakhtenbrot, B.,  Shemmer, O., Cury, Iara,
2007, ApJ, 671, 1256

Norman, C., Scoville, N., 1988, ApJ, 332, 124

Ostriker, J., Weaver, R., Yahil, A., McCray, R., 1976, ApJ, 208, L61

Padmanabhan N., et al., 2004, NewA, 9, 329

Proga, D., Ostriker, J., Kurosawa, R., 2008, ApJ, 676, 101

Reichard T.~A., Heckman T.~M., Rudnick G., 
Brinchmann J., Kauffmann G., Wild V., 2008, arXiv, arXiv:0809.3310 

Sanders D.~B., Soifer B.~T., Elias J.~H., 
Madore B.~F., Matthews K., Neugebauer G., Scoville N.~Z., 1988, ApJ, 325, 
74 

Shankar F., Crocce M., Miralda-Escude' J., 
Fosalba P., Weinberg D.~H., 2008, arXiv, arXiv:0810.4919 

Shen Y., et al., 2007, AJ, 133, 2222 

Shlosman I., Frank J., Begelman M.~C., 1989, Natur, 338, 45

Soltan A., 1982, MNRAS, 200, 115 

Springel V., Di Matteo T., Hernquist L., 2005, MNRAS, 361, 776 

Strateva I., et al., 2001, AJ, 122, 1861 

Tremaine S., et al., 2002, ApJ, 574, 740 

Ulvestad J.~S., Wilson A.~S., 1989, ApJ, 343, 659 

York D.~G., et al., 2000, AJ, 120, 1579 

Yu Q., Tremaine S., 2002, MNRAS, 335, 965

\end{document}